\newcommand{\change}{}

\documentclass{aa}
\usepackage{graphicx}
\usepackage{txfonts}

\begin{document}
%
   \title{Detection of variable Si\,II, Mn\,II and Fe\,II emission lines in the magnetic Bp star a\,Centauri\thanks{Based
on observations collected at the European Southern Observatory, Paranal, Chile
(ESO programmes 65.L-0316(A) and 073.D-0504(A)).}}

   \author{S.\ Hubrig
          \inst{1}
          \and J.F.\ Gonz\'alez
          \inst{2} 
 }

   \offprints{S. Hubrig}

   \institute{ European Southern Observatory, Casilla 19001, Santiago 19, Chile\\
              \email{shubrig@eso.org}
         \and           
            Complejo Astron\'omico El Leoncito, Casilla 467, 5400 San Juan, Argentina
}

   \date{Received ; accepted}


 \abstract
{
The nature of non-variable high-excitation emission lines detected 
in the optical spectra of normal late-B type
and chemically peculiar HgMn and PGa stars is still poorly understood.
}
{
To better understand the origin of the weak emission lines in B type stars it is 
especially important to investigate the spectra of a variety of stars to search 
for correlations between the emergence of these lines and fundamental stellar parameters.
}
{
We have acquired high resolution UVES spectra for the sharp-lined magnetic 
helium-variable star a\,Cen over the rotation period of 8.82\,d to search for the 
presence of weak emission lines.
}
{For the first time we present observational evidence 
for the appearance of variable high-excitation Si~II, Mn~II and Fe~II emission lines 
in a magnetic Bp star.  Si~II emissions are the strongest at the phase corresponding 
to the maximum strength of He~I lines.
Mn~II and Fe~II emissions vary in antiphase to the He~I lines.
A correlation is found between the probable location of Mn and Fe 
surface spots and the strength of the emission lines. 
On the basis of the currently available data it seems possible 
that the same kind of selective excitation process is working in 
the atmospheres of objects within a broad parameter space which could be defined by age, 
effective temperature, chemical composition, rotational velocity, and magnetic field. 
Neutral iron lines previously reported to appear 
broad and shallow at certain phases are not detected in our spectra, although
two of them are identified as He~I forbidden lines, showing 
maximum strength at the phase of the passage of the 
He rich region across the visible disk.
}
{}



\maketitle

\section{Introduction}
A considerable fraction of main-sequence stars of spectral
types A and B exhibits atmospheric chemical peculiarities that, apparently,
do not have an evolutionary origin, but result from the segregation of chemical
elements in the stellar outer layers under the competing actions of various
physical processes. The chemically peculiar (CP) stars are distributed into 
three main categories: the magnetic CP stars (Ap, He weak, Si and Sr-Ti, He strong),
the Am stars, and the Bp stars of HgMn and PGa peculiarity. 
{\change PGa stars are usually considered as hotter analogs of HgMn stars. The spectra 
of these stars exhibit deficient He and strongly overabundant P and Ga.} 
Although very weak magnetic 
fields have recently been detected in 
a few HgMn and one PGa star (Hubrig et al.\ \cite{hubrig06a}) this group of CP stars is usually 
regarded as non-magnetic.

The presence of weak high-excitation emission lines in the optical spectra of late-B type
chemically normal and peculiar HgMn and PGa stars has been reported 
by various authors (Sigut et al.\ \cite{sigut00}; Wahlgren \& Hubrig \cite{wahlgren00,wahlgren04};
Castelli \& Hubrig \cite{castelli04}).
The observed emission lines do not show any detectable variation and
originate from high-excitation states of a number of
elements. The spectra of P~II, Mn~II, Fe~II, Ni~II, Cu~II, Cr~II and Ti~II are
particularly rich in their number of emission lines. 
The mechanism for populating the highly-excited states is 
under investigation and the results may have a significant bearing upon our understanding
of the outermost atmospheric regions of B type stars and the formation of spectrum
anomalies generally attributed to diffusion processes. To date, explanations of the
phenomenon have been put forward in the context of non-LTE line formation 
(Sigut \cite{sigut01}) and possible fluorescence mechanisms (Wahlgren \& Hubrig \cite{wahlgren00}).

It may be an essential clue for the understanding of the origin of weak emission lines that 
they have recently been discovered in the spectrum of the very young
massive object LkH$\alpha$\,101 which illuminates the central part of the dark cloud L1482 
in the main Taurus-Aurigae cloud complex (Herbig et al.\ \cite{herbig04}). Interestingly, 
although a very early B type star was suggested for LkH$\alpha$\,101, no sign of its 
absorption spectrum could be found in the optical region. 
To better understand the nature and origin of weak emission lines in mid- to late-B 
type stars it is especially important to 
investigate the spectra of a variety of stars to search for correlations between the emergence
of these lines and effective temperature,
$\log\,g$ (as an indicator of the stellar age), membership in binary and multiple 
systems, chemical composition, peculiarity type and magnetic field.
In spite of the high resolution studies of Bp stars carried out in 
the last decade (e.g. Wade et al. (\cite{wade06}) and the numerous 
references therein), no 
detection of emission lines originating from high-excitation states of various elements has ever been 
reported for magnetic variable
helium-strong and helium-weak stars, and the qualitative and quantitative assessment of 
the spectra of B type stars seemed to link the appearance of diverse emission lines exclusively to the 
non-magnetic PGa and HgMn groups. 
Here, we present the first detection of variable Si~II, Mn~II and Fe~II emission lines
in the slowly rotating magnetic helium strong/weak variable star a\,Cen (=HD\,125823).

\begin{figure}
\centering
\includegraphics[width=0.40\textwidth]{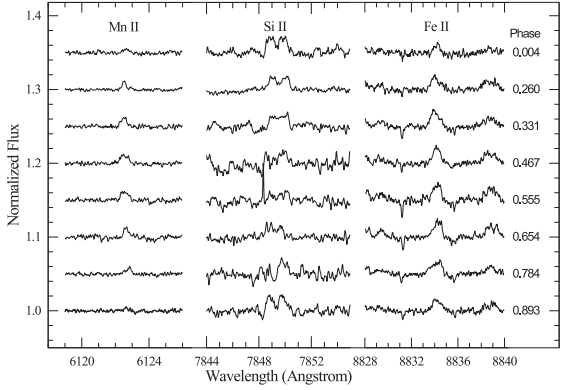}
\caption{Variations of Mn~II, Si~II and Fe~II emissions over the rotation period.
}
\label{fig:1}
\end{figure}

\begin{figure}
\centering
\includegraphics[width=0.30\textwidth]{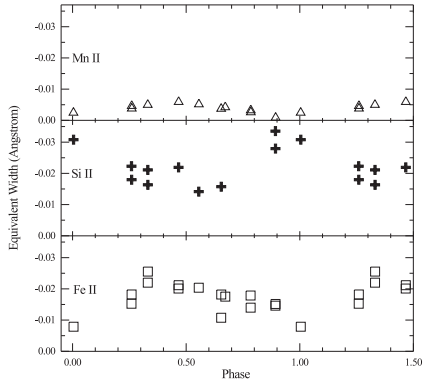}
\caption{Variations of the equivalent widths with phase.
}
\label{fig:2}
\end{figure}


\begin{figure}
\centering
\includegraphics[width=0.40\textwidth]{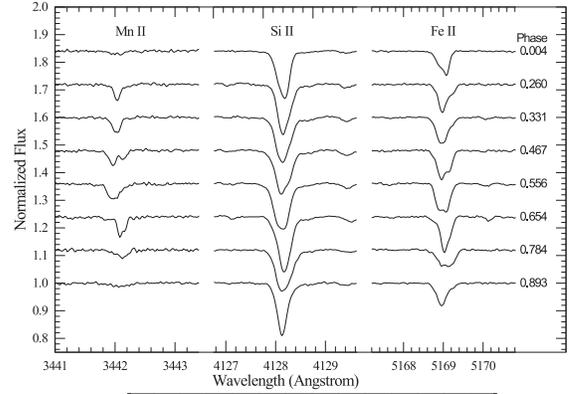}
\includegraphics[width=0.30\textwidth]{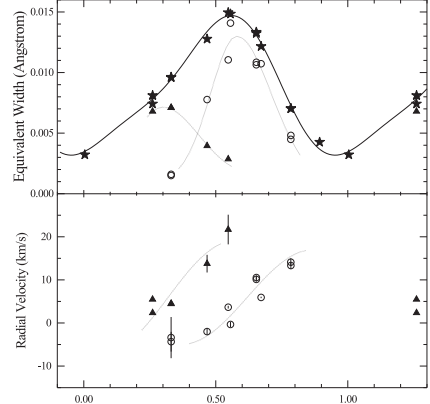}
\includegraphics[width=0.30\textwidth]{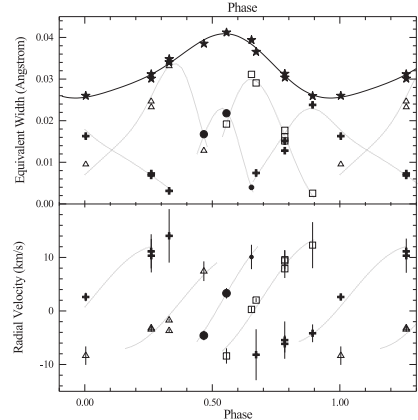}
\caption{{\sl Upper panel:} Variations of line profiles of Mn~II $\lambda$3442, 
Si~II $\lambda$4128 and Fe~II $\lambda$5169
over the rotation period. 
{\sl Middle panel:} Phase dependencies of the equivalent 
width measured by direct integration of the whole profile of the 
Mn~II line at $\lambda$3442 (filled stars), 
and of the equivalent widths measured for split line components
(filled triangles and open circles) by fitting multiple Gaussians.
The radial velocities corresponding to two 
Mn~II spots are presented in the lower half.
{\sl Lower panel:}
The equivalent widths and radial velocities for Fe~II $\lambda$5169 measured in the same 
way as for Mn~II $\lambda$3442. The surface distribution of Fe~II can be represented by 
four spots crossing the center of the 
visual disk at the phases 0.35, 0.55, 0.7 and 0.95.
}
\label{fig:4}
\end{figure}

\section{Observations and spectrum analysis}

The 5.9~M$_{\sun}$ star a\,Cen with T$_{\rm eff}$=18400\,K (Hunger \& Groote \cite{hunger99})
is a striking helium variable, ranging in helium spectral type
from He-strong B2 to He-weak B8 with a period of 8.82~d (e.g., Norris \cite{norris68}).
The magnetic field observations reported in the literature indicate that the negative
extremum coincides closely in phase with the maximum helium strength (Wolff \& Morrison \cite{wolff74}; 
Borra et al.\ \cite{borra83}).
The positive magnetic pole shows a He-deficient cap, so that the two magnetic poles
are associated with very different helium abundances. 

The observations reported here have been carried out
at the European Southern Observatory. a\,Cen was observed by us on two 
consecutive nights in May 2000 with the echelle spectrograph FEROS on the 1.52\,m
telescope at La Silla at a resolving power of 48,000. Rather unexpectedly, we 
detected in the FEROS spectra a few very weak emission lines of Mn~II and Fe~II. Some other lines 
offered suspicions that they could also be in emission, but our data 
were of too low spectral resolution and too scant in phase to 
be definitive. In May 2004 we were able to obtain 
much higher resolution and higher signal-to-noise ratio UVES spectra on nine consecutive nights 
to cover the rotation period of 8.82\,d. 
We used the UVES DIC1 and DIC2 standard settings to cover the spectral 
range from 3030\,\AA{} to 10,000\,\AA{}.
The slit width was set to $0\farcs{}3$ for the red arm, 
corresponding to a resolving power of 
$\lambda{}/\Delta{}\lambda{} \approx 110,000$. For the blue arm, we used 
a slit width of $0\farcs{}4$ to achieve a resolving power of 
$\approx 80,000$.
The spectra were reduced by the UVES pipeline Data Reduction Software (version 2.5; Ballester 
et al.\ 2000) and using standard IRAF routines. 
The signal-to-noise ratio of the obtained UVES spectra is very high, 
ranging from 350 in the near UV to 600 in the visual region.
The measured $v\,\sin i$ value for a\,Cen is only about 15~km$\,$s$^{-1}$ and high
signal-to-noise ratio and high spectral resolution are critical to precisely define line profile
shapes. 


\begin{figure*}
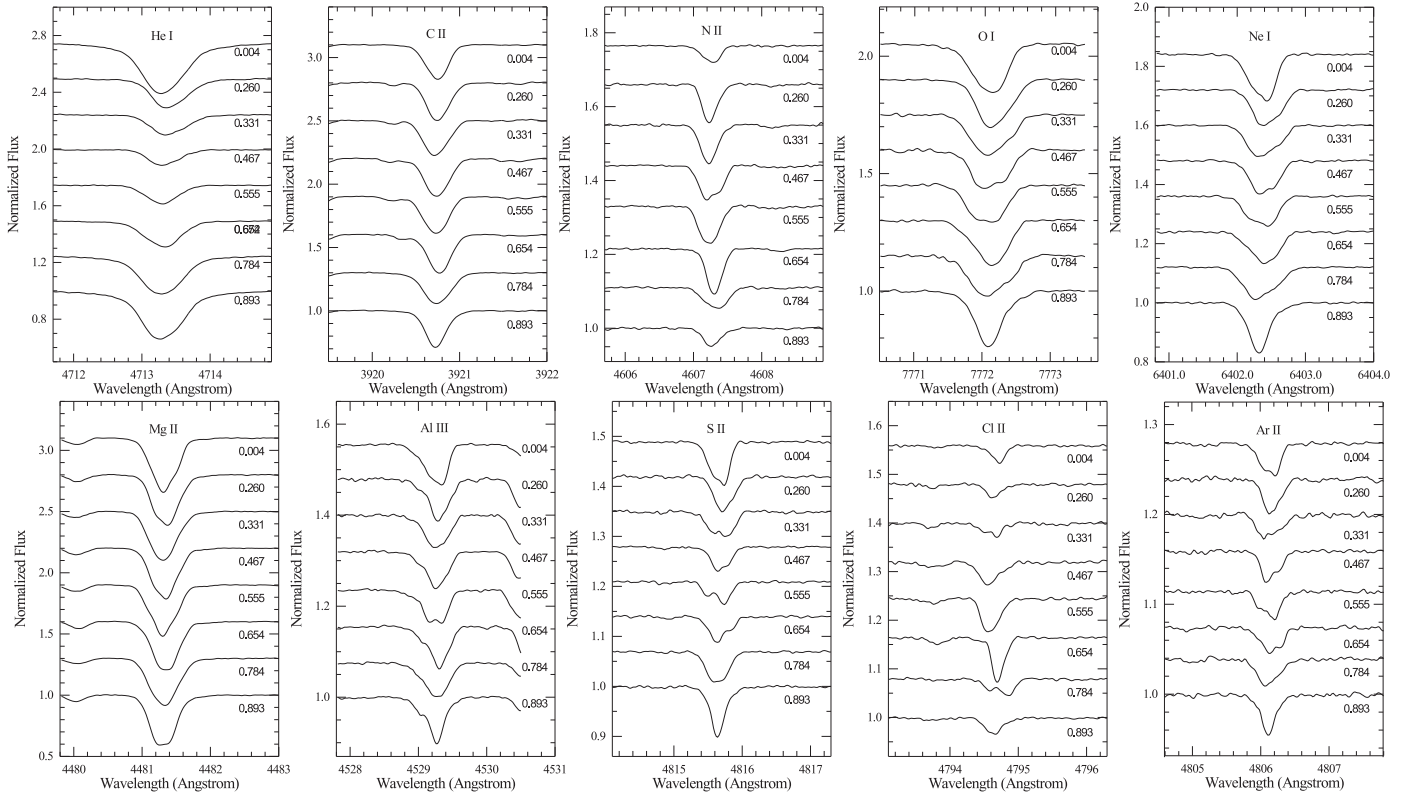

\centering
\includegraphics[width=\textwidth]{6738f4a.eps}
\includegraphics[width=0.99\textwidth]{6738f4b.eps}
\caption{Variations of line profiles of various elements phased with the rotation period.
}
\label{fig:7}
\end{figure*}

\begin{figure*}
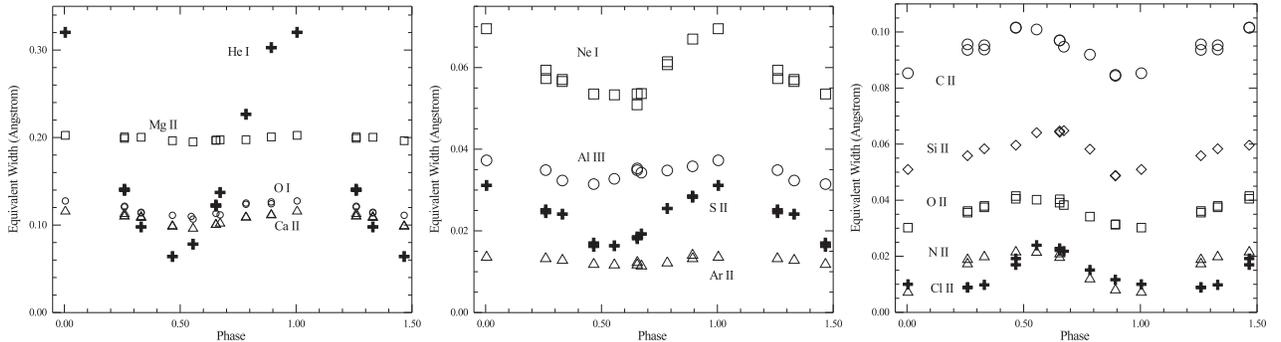

\centering
\includegraphics[width=0.3\textwidth]{6738f5a.eps}
\includegraphics[width=0.3\textwidth]{6738f5b.eps}
\includegraphics[width=0.302\textwidth]{6738f5c.eps}
\caption{Variations of the equivalent width of various elements phased with the 
rotation period.
}
\label{fig:8}
\end{figure*}

\begin{table}
\centering
\caption{Emission lines detected in the spectrum of a\,Cen.}
\label{tab:1}
\begin{tabular}{lcllc}
\multicolumn{1}{c}{$\lambda$ [\AA]} &
\multicolumn{1}{c}{Identification} & &
\multicolumn{1}{c}{$\lambda$ [\AA]} &
\multicolumn{1}{c}{Identification} \\
\cline{1-2}\cline{4-5}
6122.43   &    Mn~II & & 8769.16    &    Mn~II \\
6239.80    &    unidentified          & & 8774.57    &    unidentified          \\
7513.14    &    Fe~II & & 8813.33    &    Fe~II \\
7848.75    &    Si~II & & 8825.54    &    unidentified          \\
7849.76    &    Si~II & & 8834.01    &    Fe~II \\
8490.05    &    Fe~II & & 8910.39    &    unidentified          \\
8767.63    &    unidentified          & &            &  \\
\end{tabular}
\end{table}

In our spectra, weak emission lines are clearly seen at the locations of several 
Si~II, Mn~II and Fe~II lines. 
A number of He-rich and He-weak magnetic stars appear to possess comparatively dense magnetospheres
which are observed via Balmer or helium emission 
line variations, or anomalously strong and variable UV resonance lines (e.g. Shore et al. \cite{shore04}).
We note that no emissions are detected in the H~I and He~I lines, which means that we do not detect 
any evidence for an extended atmosphere..
The list of observed emission lines 
is presented in Table~\ref{tab:1}.
The detected emission lines are definitely variable over the star's rotation period.
The Si~II emission reaches the largest intensity at the phase corresponding to the maximum strength
of the helium lines. Mn~II and Fe~II emissions vary in antiphase to the He~I lines.
This is the first time that the variability of emission lines is detected in B type stars.
The variations of the line profiles of the emission lines of Si~II,
Mn~II and Fe~II are 
presented in Fig.~\ref{fig:1}. In Fig.~\ref{fig:2} we show the variation
of their equivalent widths over the rotation period.
Although the Mn~II line at $\lambda$6122.4 of multiplet 13
(transition 4d\,$^5$D--4f\,$^5$F) is observed in emission, 
the strongest line of multiplet 11 at 
$\lambda$5302.32, with a comparable upper excitation 
potential, appears in absorption. This observation suggests that a selective excitation process 
might be at work populating the multiplet 13 upper term. The same anomaly has been observed 
in $\eta$\,Car by Johansson et al.\ (\cite{johansson95}) and in HgMn and normal late-B type stars by
Wahlgren \& Hubrig (\cite{wahlgren00}).
Among the detected emission lines only the unidentified line at $\lambda$6239.8 shows a P\,Cygni profile. 
The character of the intensity variations of the Mn~II and Fe~II absorption lines in the spectra of a\,Cen 
is very similar to that of the emission lines, reaching the maximum intensity at the phase corresponding 
to the minimum strength of the He~I lines. However, Si~II emissions vary in antiphase to the Si~II absorption 
lines. Due to the inhomogeneous surface distribution of Mn~II and Fe~II the 
absorption line profiles of these elements appear split at phases 
0.47-- 0.78, indicating that two or more areas of enhanced Mn and Fe are crossing the visible 
disk  at these phases (Fig.~\ref{fig:4}, upper panel). The Si~II lines do not appear 
split, but the phase variation of the Si~II 
line profile at $\lambda$4128 indicates that Si~II is also
distributed inhomogeneously over the stellar surface. 
The measurements of equivalent width and radial velocity of the Mn~II line at
$\lambda$3442 and the Fe~II line at $\lambda$5169 are presented in 
Fig.~\ref{fig:4} in the middle and lower panels.
At the phases of splitting, the radial velocities and equivalent widths of the split components 
have been measured by fitting multiple Gaussians. The variations of 
the Mn~II line at  $\lambda$3442
can be presented by two spots of enhanced Mn~II  crossing the center of the visual 
disk at phases 0.3 and 0.6.
Fe~II is likely enhanced in four spots crossing the center of the 
visual disk at phases 0.35, 0.55, 0.7 and 0.95. 



\begin{figure}
\centering
\includegraphics[width=0.30\textwidth]{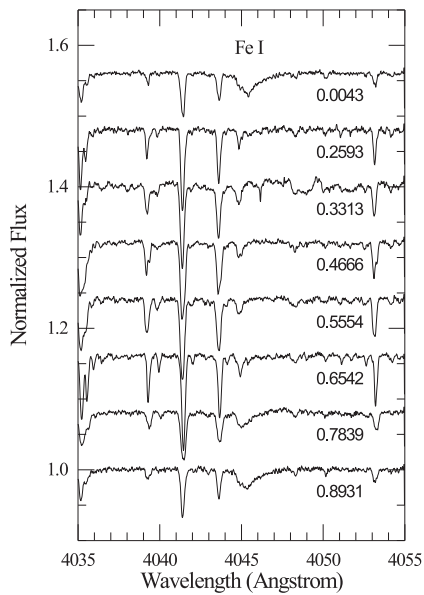}
\caption{The behaviour of the $\lambda$4045 line over the rotation period.
}
\label{fig:9}
\end{figure}

Inspecting the behaviour of the spectral lines of other elements at different rotation phases we conclude that 
all elements show a spectral variability. The equivalent widths of lines of C~II, N~II, O~II, 
Si~II, Cl~II, Mn~II and Fe~II
appear variable, showing their maximum in the phase corresponding to the minimum strength
of He~I, O~I, Ne~I, Mg~II, Al~III, S~II, Ar~II and Ca~II. 
We present the behaviour of the line profiles of various elements over the 
rotation period in Fig.~\ref{fig:7}.
The variations of the equivalent widths are shown in Fig.~\ref{fig:8}.
It is interesting that for some elements
the equivalent widths of neutral and ionised lines vary in antiphase. This phenomenon has already been
mentioned in previous studies by
Underhill et al.\ (\cite{underhill75}) and Norris \& Baschek (\cite{norris72}).
An example of this opposite
behaviour of equivalent widths of O~I and O~II is presented in Fig.~\ref{fig:8}.

The appearance of broad shallow Fe~I lines in the spectrum of a\,Cen at certain phases has 
previously been reported by Underhill \& Klinglesmith (\cite{underhill73}). However, the existence of these lines
could not be confirmed by our observations. It is interesting that at least two of the observed Fe~I lines appear  
at the wavelength positions of He~I forbidden lines, at $\lambda\lambda$4045.2 and 4383.5, which become 
significantly stronger at phases of maximum strength of the helium lines (Fig.~\ref{fig:9}).

Previous studies of a\,Cen suggest that this star is seen from an inclination $i\gtrsim30^\circ$
and has an angle $\beta$ between the magnetic and rotation axes of nearly $90^\circ$ 
(Borra et al.\ \cite{borra83}).
The line profile variations as well as the variations of the equivalent width and 
of the radial velocity of the He lines observed in our spectra support the models of a surface 
distribution with He-rich and He-weak hemispheres (Borra et al.\ \cite{borra83}; Vet\"o et al.\ \cite{veto91}). 
It is remarkable that the
He~I lines show some hint of splitting around the rotational phase 0.5. Such an appearance can be explained by a 
He-weak spot passing the center of the visible disk at this phase.
The lines of 
N~II, O~I, Al~III, Cl~II, Ar~II, S~II, N~II, Ca~II, Mn~II and Fe~II
show a similar splitting at the same phase as the He~I lines.
This behaviour could possibly be explained by an inhomogeneous surface distribution of these 
elements involving several 
spots in the transition region between the He-rich and He-weak spots, or by a continuous ring 
around the positive magnetic pole with a He-deficient cap. 
The detailed geometry of the surface distribution of He and the other elements still has to be determined.
However, since the measured $v\,\sin i$ value for a\,Cen is only 
about 15\,km/s, it is rather hopeless to apply the standard Doppler Imaging technique. As a future task we
plan to use 
the so-called direct Doppler Imaging method allowing to compute spectral line profiles from test 
images created by varying the local element abundances (Savanov \& Strassmeier \cite{savanov05};
Hubrig et al.\ \cite{hubrig06b}).

\section{Discussion}\label{sec3}

{\change The fact that large-scale organized magnetic 
fields of Ap and Bp stars are more readily observable than
those of any other type of non-degenerate stars makes them a privileged
laboratory for the study of phenomena related to stellar magnetism.
The rather bright slowly rotating star a\,Cen combines a well measured magnetic field with 
remarkable spectrum variations and thus deserves further detailed studies to clarify 
the relationship between the appearance of weak high-excitation emissions, vertical and horizontal
abundance gradients and magnetic field geometry.}
The origin of weak high-excitation emission lines in various B type 
stars is still rather poorly understood.
Theoretical modelling involving non-LTE photospheric models and a simplistic two level
inverse stratification of abundance has been tested for certain multiplets of Mn\,II 
(Sigut \cite{sigut01}).
This modelling requires a photospheric origin for the emission and made specific predictions 
regarding the strength of Mn\,II emission lines. A second approach has been put forward by 
Wahlgren \& Hubrig (\cite{wahlgren00}) that postulates a connection between atomic 
metastable states with absorption lines found in the vicinity of hydrogen Ly$\alpha$ and  Ly$\beta$ 
and emission lines observed in the red spectral region.

The importance of detecting emission lines in different groups of B type stars 
lies in its possible relevance to diffusion theory and the development of spectrum 
variability. It is the first time that emission lines and their variability are definitely 
identified in the spectra of a magnetic star. 
Although a number of members of the PGa or HgMn groups
occur at the same effective temperature as a\,Cen,
this star has never been reported to show the distinctive abundance anomalies of 
either the PGa or HgMn stars. 
This fact implies that a\,Cen presents a particularly interesting case for forging 
connections between different chemically peculiar star classes. 
The detection of emission lines
in this star
alters our perception of distinctions between the various chemically peculiar star classes and
should provide a basis for a more unified picture of the physical processes that may be 
responsible for the observed spectral anomalies. 

The non-detection 
of weak emissions in other magnetic Bp stars could be related to the 
rather large $v\,\sin i$ values of the stars previously studied by other authors. In our study of emission 
lines in the spectra of late-B type stars (Wahlgren \& Hubrig \cite{wahlgren00}) we noted that 
the emission appears 
to be formed near the photosphere, since the emission line widths correspond to those of the underlying 
rotationally broadened absorption lines.
Thus it is possible that other magnetic Bp star atmospheres 
produce emission lines that are difficult to detect in broad-lined rapidly rotating stars.

Previous studies of stars with emission lines have shown that the determination of 
element abundances is especially important to establish correlations between the presence
of emission and abundance patterns. For example, the HgMn stars do not display Mn\,II emissions
if the abundance enhancement of Mn is greater than about a factor of 10 above the solar value
(Wahlgren \& Hubrig \cite{wahlgren00}). The Cr\,II emission appears to be correlated with 
the Cr abundance in the sense that the most Cr-enhanced stars display the largest emission equivalent
widths and they show the most developed Cr\,II emission spectra. 
The  study of the dependence of the emission line variations on the distribution of the 
surface abundance over the stellar surface in a\,Cen must therefore give us clues to the
structure of the stellar atmosphere and whether diffusion and/or stellar winds may be present.

It has been shown that the Mn~II and Fe~II high-excitation emissions 
also appear in the spectra of objects of completely different evolutionary states.
As mentioned in the introduction, apart from the detection of these lines in the spectra of main-sequence B type stars, 
they have also been 
discovered in the spectra of the luminous blue variable $\eta$\,Car and of the young massive star
LkH$\alpha$\,101 in a massive star forming region. 
The Mn~II and Fe~II emissions, which have been identified 
in a\,Cen, have also been detected in our UVES spectrum of the Population II star Feige\,86 
with the atmospheric parameters T$_{\rm eff}$=16430\,K and $\log g$=4.20
(Castelli et al., in preparation). 
Right now, on the basis of the currently available data, it seems possible that the same kind of selective 
excitation process is working in 
the atmospheres of objects within a broad parameter space defined by age, effective temperature, 
chemical composition, rotation velocity, and magnetic field.

\end{document}